\documentclass[12pt]{article}
\usepackage{graphicx}
\usepackage{lscape}
\usepackage{citesort}
\usepackage{amssymb}
\usepackage{appendix}
\usepackage{multirow}

\begin{document}
\def\qq{\langle \bar q q \rangle}
\def\uu{\langle \bar u u \rangle}
\def\dd{\langle \bar d d \rangle}
\def\sp{\langle \bar s s \rangle}
\def\GG{\langle g_s^2 G^2 \rangle}
\def\Tr{\mbox{Tr}}
\def\figt#1#2#3{
        \begin{figure}
        $\left. \right.$
        \vspace*{-2cm}
        \begin{center}
        \includegraphics[width=10cm]{#1}
        \end{center}
        \vspace*{-0.2cm}
        \caption{#3}
        \label{#2}
        \end{figure}
    }

\def\figb#1#2#3{
        \begin{figure}
        $\left. \right.$
        \vspace*{-1cm}
        \begin{center}
        \includegraphics[width=10cm]{#1}
        \end{center}
        \vspace*{-0.2cm}
        \caption{#3}
        \label{#2}
        \end{figure}
                }

\def\ds{\displaystyle}
\def\beq{\begin{equation}}
\def\eeq{\end{equation}}
\def\bea{\begin{eqnarray}}
\def\eea{\end{eqnarray}}
\def\beeq{\begin{eqnarray}}
\def\eeeq{\end{eqnarray}}
\def\ve{\vert}
\def\vel{\left|}
\def\ver{\right|}
\def\nnb{\nonumber}
\def\ga{\left(}
\def\dr{\right)}
\def\aga{\left\{}
\def\adr{\right\}}
\def\lla{\left<}
\def\rra{\right>}
\def\rar{\rightarrow}
\def\lrar{\leftrightarrow}
\def\nnb{\nonumber}
\def\la{\langle}
\def\ra{\rangle}
\def\ba{\begin{array}}
\def\ea{\end{array}}
\def\tr{\mbox{Tr}}
\def\ssp{{\Sigma^{*+}}}
\def\sso{{\Sigma^{*0}}}
\def\ssm{{\Sigma^{*-}}}
\def\xis0{{\Xi^{*0}}}
\def\xism{{\Xi^{*-}}}
\def\qs{\la \bar s s \ra}
\def\qu{\la \bar u u \ra}
\def\qd{\la \bar d d \ra}
\def\qq{\la \bar q q \ra}
\def\gGgG{\la g^2 G^2 \ra}
\def\q{\gamma_5 \not\!q}
\def\x{\gamma_5 \not\!x}
\def\g5{\gamma_5}
\def\sb{S_Q^{cf}}
\def\sd{S_d^{be}}
\def\su{S_u^{ad}}
\def\sbp{{S}_Q^{'cf}}
\def\sdp{{S}_d^{'be}}
\def\sup{{S}_u^{'ad}}
\def\ssp{{S}_s^{'??}}

\def\sig{\sigma_{\mu \nu} \gamma_5 p^\mu q^\nu}
\def\fo{f_0(\frac{s_0}{M^2})}
\def\ffi{f_1(\frac{s_0}{M^2})}
\def\fii{f_2(\frac{s_0}{M^2})}
\def\O{{\cal O}}
\def\sl{{\Sigma^0 \Lambda}}
\def\es{\!\!\! &=& \!\!\!}
\def\ap{\!\!\! &\approx& \!\!\!}
\def\ar{&+& \!\!\!}
\def\ek{&-& \!\!\!}
\def\kek{\!\!\!&-& \!\!\!}
\def\cp{&\times& \!\!\!}
\def\se{\!\!\! &\simeq& \!\!\!}
\def\eqv{&\equiv& \!\!\!}
\def\kpm{&\pm& \!\!\!}
\def\kmp{&\mp& \!\!\!}
\def\mcdot{\!\cdot\!}
\def\erar{&\rightarrow&}


\def\simlt{\stackrel{<}{{}_\sim}}
\def\simgt{\stackrel{>}{{}_\sim}}

\def\olra{\stackrel{\leftrightarrow}}
\def\ola{\stackrel{\leftarrow}}
\def\ora{\stackrel{\rightarrow}}


\title{
         {\Large
                 {\bf
Magnetic dipole moment of the light tensor mesons in light cone QCD sum
rules
                 }
         }
      }

\author{\vspace{1cm}\\
{\small T. M. Aliev \thanks {e-mail:
taliev@metu.edu.tr}~\footnote{permanent address:Institute of
Physics,Baku,Azerbaijan}\,\,, K. Azizi \thanks
{e-mail:kazizi@dogus.edu.tr}\,\,, M. Savc{\i} \thanks
{e-mail: savci@metu.edu.tr}} \\
{\small $^{\dag,\S}$Physics Department, Middle East Technical University,
06531 Ankara, Turkey}\\
{\small$^{\ddag}$ Physics Division,  Faculty of Arts and Sciences, Do\u gu\c s University,}\\
{\small Ac{\i}badem-Kad{\i}k\"oy,  34722 Istanbul, Turkey} }
\date{}

\begin{titlepage}
\maketitle
\thispagestyle{empty}

\begin{abstract}
The magnetic dipole moments of the light tensor mesons $f_2$,
$a_2$ and strange $K_2^{\ast 0} (1430)$ tensor meson
are calculated in the framework of the light cone QCD sum rules.
It is observed that the values of the magnetic dipole moment for
the charged tensor particles are considerably different from zero.
These values are very close to zero for the light  neutral $f_2$ and
$a_2$
 tensor mesons, while it has a small nonzero value for the
neutral strange $K_2^{\ast 0} (1430)$ tensor meson.
\end{abstract}

\vspace{1cm}
~~~PACS number(s): 11.55.Hx, 13.40.Em, 13.40.Gp
\end{titlepage}

\section{Introduction}

Investigation of the properties of the hadrons such as
electromagnetic form factors and multipole moments can shed light in
understanding their internal structure, as well as their geometric shape.
The electromagnetic properties of non-tensor mesons, as well as light and
heavy baryons have been widely discussed in the literature. The
electromagnetic form factors of the pseudoscalar $\pi$ mesons are investigated
extensively (see \cite{Rte01}--\cite{Rte06} and references therein). Form
factors of the vector mesons are studied in \cite{Rte07}--\cite{Rte11}, as
well as in lattice QCD \cite{Rte06,Rte07,Rte12}. The magnetic and
quadrupole moments of light vector and axial--vector mesons in light cone
QCD sum rules (LCSR) are investigated in \cite{Rte13} (for more about LCSR,
see \cite{Rte14} and \cite{Rte15}). It should be noted that
static properties of baryons within the LCSR method are
studied in \cite{Rte16} and \cite{Rte17}.
However these properties of tensor mesons have received less interest and further
detailed analysis is needed in this respect. So far, only the mass and decay
constants of the light, unflavored tensor mesons within QCD sum rules is
studied \cite{Rte18}. Recently, similar calculation is extended to cover
the strange tensor mesons \cite{Rte19}. In the present
work, we calculate the magnetic dipole moment of the light tensor mesons
in the LCSR method.

The outline of the paper is as follows: In section 2, the sum rules for the
the magnetic dipole moments of the light tensor mesons are obtained in the
framework of the LCSR method. Section 3 is devoted to the
numerical analysis of the magnetic dipole moment and discussion.

\section{Light cone QCD sum rules for the magnetic moment of the light
tensor mesons}

In order to obtain the sum rules for the magnetic dipole moment, we
consider the following correlation function:
\bea
\label{ete01}
\Pi_{\mu\nu\alpha\beta} = i \int d^4x \,  e^{ipx }
\lla 0 \vel {\cal T} \Big\{ j_{\mu\nu} (x) \bar{j}_{\alpha\beta} (0) \Big\}
\ver 0 \rra_\gamma~,
\eea
where, ${\cal T}$ is the time ordering, $j_{\mu\nu}$ and
$j_{\alpha\beta}$ are the interpolating currents
corresponding to the initial and final states, with $p$ being the momentum of
the final state and $q$ is the momentum transfer, and $\gamma$ is the external
electromagnetic field, respectively. The interpolating current $j_{\mu\nu}$ of
the ground state tensor mesons is given as:
\bea
\label{ete02}
j_{\mu\nu} = {i\over 2} \Big[ \bar{q}_1(x) \gamma_\mu
\olra{\cal D}_\nu (x)
q_2(x) + \bar{q}_1(x) \gamma_\nu \olra{\cal D}_\mu (x)
q_2(x) \Big]~,
\eea
where $\olra{\cal D}_\mu (x)$ represents the derivative with
respect to $x_\mu$ acting on the right and left sides simultaneously, which
is defined as:
\bea
\label{ete03}
\olra{\cal D}_\mu (x) = {1\over 2} \Big[
\ora{\cal D}_\mu (x) -
\ola{\cal D}_\mu (x) \Big]~.
\eea
The covariant derivative defined in Eq. (\ref{ete03}) can be written in
terms of the normal derivative and the external (vacuum) gluon fields as
follows:
\bea
\label{ete04}
\ora{\cal D}_\mu (x) \es \ora{\partial}_\mu (x) - i
{g\over 2} \lambda^a A_\mu^a (x) ~, \nnb \\
\ola{\cal D}_\mu (x) \es \ola{\partial}_\mu (x) + i
{g\over 2} \lambda^a A_\mu^a (x) ~,
\eea
where $\lambda^a$ are the Gell--Mann matrices.

In further analysis we will use the Fock--Schwinger gauge. The main advantage
of this gauge is that  the external field is expressed in terms of the
field strength tensor, i.e.,
in the Fock--Schwinger gauge, where the condition $x^\mu A_\mu^a (x)=0$
is imposed, we have:
\bea
\label{ete05}
A_\mu^a (x)= \int_0^1 d \alpha \, \alpha x_\beta G_{\beta\mu}^a (\alpha x)~.
\eea
It follows from Eq. (\ref{ete02}) that, the current of the tensor mesons contains
derivatives, and therefore we take into account the initial state at  point  $y$, and then
after carrying out calculations, we set it to zero.

We can now proceed to obtain the sum rules for the magnetic dipole moment of
the light tensor mesons. This sum rules are obtained from the following
three steps.

\begin{itemize}
\item
The correlation function in Eq. (\ref{ete01}) is calculated in
hadronic language, so--called the physical or phenomenological side. In
order to obtain the expression from the physical side, the correlation function
is saturated with a tower of tensor mesons
having the same quantum numbers as the interpolating currents.

\item
The aforementioned correlation function is calculated in quark--gluon
language called the theoretical or QCD side. In this representation, the
correlator is calculated in deep Euclidean region with the help of the
operator product expansion (OPE), where the short and long distance
quark--gluon interactions are factored out. The short distance effects are
calculated using the perturbation theory, whereas the long distance
effects are parametrized in terms of the photon distribution amplitudes (DA's).

\item
The two representations of the correlation function in the above steps are
equated through the dispersion relation. To suppress the contribution of the
higher states and continuum, the Borel transformation and continuum
subtraction are applied to both sides of the equality.

\end{itemize}

Let us first calculate the physical side of the  correlation function. Inserting the complete sets
of mesonic states into correlation function in Eq. (\ref{ete01}), and
isolating the ground state, we obtain:

\bea
\label{ete06}
\Pi_{\mu\nu\alpha\beta} = i{\lla 0 \vel j_{\mu\nu} \ver T(p,\varepsilon) \rra
\over p^2 - m_T^2} \lla T(p,\varepsilon) \vel T(p+q,\varepsilon) \right.
\rra_\gamma {\lla T(p+q,\varepsilon ) \vel j_{\alpha\beta} \ver 0 \rra \over
(p+q)^2 - m_T^2 } + \cdots ~,
\eea
where $T(p,\varepsilon)$ denotes the light tensor mesons with momentum $p$
and  polarization tensor $\varepsilon$. Here, $\cdots$ represents the
contribution of the higher states and the continuum. The vacuum to the
mesonic state of the interpolating current is parametrized in terms of the
decay constant and the polarization tensor  as:

\bea
\label{ete07}
\lla 0 \vel j_{\mu\nu} \ver T(p,\varepsilon) \rra = m_T^3 g_T
\varepsilon_{\mu\nu}~.
\eea

The transition matrix element $\lla T(p,\varepsilon) \ve T(p+q,\varepsilon)
\rra_\gamma$ in terms of the form factors is determined as follows:
\bea
\label{ete08}
\lla T(p,\varepsilon) \ve T(p+q,\varepsilon) \rra_\gamma \es
\varepsilon_{\alpha^\prime\beta^\prime}^\ast (p) \Bigg\{ 2 (\varepsilon^{\prime}. p)
\Bigg[ g^{\alpha^\prime\rho} g^{\beta^\prime\sigma} F_1
- g^{\beta^\prime\sigma} {q^{\alpha^\prime }q^\rho \over 2 m_T^2} F_3 + {q^{\alpha^\prime} q^\rho
\over 2 m_T^2}\, {q^{\beta'} q^\sigma \over 2 m_T^2} F_5 \Bigg] \nnb \\
\ar (\varepsilon^{\prime\sigma} q^{\beta'} - \varepsilon^{\prime\beta'} q^\sigma)
\Bigg[ g^{\alpha'\rho} F_2 - {q^{\alpha'} q^\rho \over 2 m_T^2} F_4 \Bigg]
\Bigg\} \varepsilon_{\rho\sigma} (p+q)~,
\eea
where $F_i(q^2)$ are the form factors and $\varepsilon^{\prime}$ is the photon
polarization vector.

However, in the experiments it is more convenient to use the set of form
factors which correspond to a definite multipole in a given reference frame.
Relations between the two sets of form factors for the arbitrary integer spin
(as well as arbitrary half--integer) case are obtained in \cite{Rte20}. In
our case, i.e., for the real photon case $q^2=0$, these relations are rather
simple and are as follows:
\bea
\label{ete09}
F_1(0) \es G_{E_0}(0)~, \nnb \\
F_2(0) \es G_{M_1}(0)~, \nnb \\
F_3(0) \es - 2 G_{E_0}(0) +  G_{E_2}(0) + G_{M_1}(0)~, \nnb \\
F_4(0) \es - G_{M_1}(0) + G_{M_3}(0)~, \nnb \\
F_5(0) \es G_{E_0}(0) -  G_{E_2}(0) - G_{M_1}(0)  + G_{E_4}(0) +
G_{M_3}(0)~,
\eea
where $G_{E_\ell}(0)$ and $G_{M_\ell}(0)$ are the electric and magnetic
multipoles.

Using Eqs. (\ref{ete08}) and (\ref{ete09}), the transition matrix element
$\lla T(p,\varepsilon) \ve T(p+q,\varepsilon) \rra_\gamma$ can be written in
terms of the electric and magnetic multipoles as:

\bea
\label{ete10}
\lla T(p,\varepsilon)\vel T(p+q,\varepsilon) \right.\rra_\gamma \es
\varepsilon_{\alpha^\prime\beta^\prime}^\ast (p) \Bigg\{ 2 (\varepsilon^{\prime} \cdot p)
\Bigg[ g^{\alpha^\prime\rho} g^{\beta^\prime\sigma} G_{E_0} -
{q^{\alpha^\prime} q^\rho \over 2 m_T^2} g^{\beta^\prime\sigma}
(-2 G_{E_0} + G_{E_2} + G_{M_1} ) \nnb \\
\ar {q^{\alpha^\prime} q^\rho \over 2 m_T^2} {q^{\beta^\prime} q^\sigma
\over 2 m_T^2} (  G_{E_0} - G_{E_2} - G_{M_1} + G_{E_4} + G_{M_3} )
\Bigg] \nnb \\
\ar (\varepsilon^{\prime \sigma} q^{\beta^\prime} -
\varepsilon^{\prime\beta^\prime} q^\sigma ) \Bigg[ g^{\alpha^\prime\rho} G_{M_1} -
{q^{\alpha^\prime} q^\rho \over 2 m_T^2} ( - G_{M_1} + G_{M_3} ) \Bigg]
\Bigg\} \varepsilon_{\rho\sigma} (p+q)~.\nnb\\
\eea
Substituting Eqs. (\ref{ete07}) and (\ref{ete10}) into Eq. (\ref{ete06}) and
performing summation over the polarizations of spin--2 particles using the relation,

\bea
\label{ete11}
\varepsilon_{\mu\nu} (p) \varepsilon_{\alpha\beta}^\ast (p) \es
{1\over 2} \Bigg( - g_{\mu\alpha} + {p_\mu p_\alpha \over m_T^2} \Bigg)
\Bigg( - g_{\nu\beta} + {p_\nu p_\beta \over m_T^2} \Bigg) \nnb \\
\ar {1\over 2} \Bigg( - g_{\mu\beta} + {p_\mu p_\beta \over m_T^2} \Bigg)
\Bigg( - g_{\nu\alpha} + {p_\nu p_\alpha \over m_T^2} \Bigg) \nnb \\
\ek {1\over 3} \Bigg( - g_{\mu\nu} + {p_\mu p_\nu \over m_T^2} \Bigg)
\Bigg( - g_{\alpha\beta} + {p_\alpha p_\beta \over m_T^2} \Bigg)~,
\eea
one gets the final expression of the correlation function on
the physical side. Obviously, the correlation function contains many
independent structures encountered, and among these structures
we can choose any independent one for determination of the multipole form
factors. In the present work, we restrict ourselves to calculate the magnetic
dipole form factor only
and for this aim the structure $(\varepsilon^{\prime\beta} q^\nu -\varepsilon^{\prime\nu} q^\beta) g^{\mu\alpha}$
is chosen. The choice of this structure is dictated by the fact that it
does not get contribution from the contact terms (for a discussion about
contact terms see \cite{Rte21}). Separating out the coefficient of this
structure we get the final expression for the correlation function in physical side in terms of the magnetic dipole moment as:
\bea
\label{ete12}
\Pi = i{m_T^6 g_T^2 \over (p^2-m_T^2) [(p+q)^2 - m_T^2]}
\Bigg\{ {1\over 4} G_{M_1} +
\mbox{\rm other structures} \Bigg\} + \cdots ~.
\eea

On the QCD side, the correlation function is calculated
in deep Euclidean region where $p^2 \rar -\infty$ and $(p+q)^2 \rar
-\infty$, via the OPE. After contracting out all quark pairs, we obtain the following
representation of the correlation function on the theoretical side:

\bea
\label{ete13}
\Pi_{\mu\nu\alpha\beta} \es
{-i \over 16} \int e^{ip\cdot x}  d^4x\,
\lla \gamma \vel \Bigg\{
S_{q_1} (y-x) \gamma_\mu \Big[
 \ora{\partial}_\nu (x) \ora{\partial}_\beta (y) \right. \right.\nnb \\
\ek \left. \left.
 \ora{\partial}_\nu (x) \ola{\partial}_\beta (y)
-
 \ola{\partial}_\nu (x) \ora{\partial}_\beta (y)  +
 \ola{\partial}_\nu (x) \ola{\partial}_\beta (y) \Big]
S_{q_2} (x-y) \gamma_\alpha \Bigg\}\ver 0 \rra  \nnb \\
\ar
\{ \beta \leftrightarrow \alpha \} + \{ \nu \leftrightarrow \mu \} +
\{ \beta \leftrightarrow \alpha,~ \nu \leftrightarrow \mu \}~.
\eea

In order to proceed with the analysis, we need to know the explicit
expressions for the light quark propagator. The light quark propagator in
the external field is calculated in \cite{Rte03,Rte04}, which has the form:

\bea
\label{ete14}
S_q(x-y) \es S^{free} (x-y) - {\langle qq\rangle \over 12} \Bigg[1 -i {m_q \over
4} (\not\!{x}-\not\!{y}) \Bigg] - {(x-y)^2 \over 192} m_0^2 \langle qq\rangle
\Bigg[1 -i {m_q \over 6} (\not\!{x}-\not\!{y}) \Bigg] \nnb \\
\ek i g_s \int_0^1 du \Bigg\{
{\not\!x - \not\!{y} \over 16 \pi^2 (x-y)^2} G_{\mu\nu}(u(x-y)) \sigma^{\mu\nu}
- u (x^\mu-y^\mu) G_{\mu\nu}(u(x-y)) \gamma^\nu \nnb \\
\cp {i \over 4 \pi^2 (x-y)^2}
- i {m_q \over 32 \pi^2} G_{\mu\nu}(u(x-y)) \sigma^{\mu\nu} \Bigg[
\ln\left( -{(x-y)^2
\Lambda^2 \over 4} + 2 \gamma_E \right) \Bigg] \Bigg\}~,
\eea
where $\Lambda$ is the scale parameter and we choose it as a factorization
scale, i.e., $\Lambda=(0.5-1)~GeV$ \cite{Rte22}. Here, we would like to made the following remark about the expression of the quark propagator. The complete light cone expansion of the light quark propagator is
 given in \cite{balitskyb} and it gets contributions from nonlocal $\bar qGq$, $\bar qG^2q$ and $\bar qq\bar qq$ operators, where $G_{\mu\nu}$ is the gluon field strength tensor. In the present work, we take into 
account the operators with only one gluon field and neglect the contributions coming from the $\bar qG^2q$ and $\bar qq\bar qq$ operators. Formally, ignoring from these terms can be justified on the basis of an 
expansion in conformal spin \cite{veladimirb}.

The free quark operator is given as:
\bea
\label{ete15}
S^{free} (x-y) = {i \not\!x-\not\!y \over 2 \pi^2 (x-y)^4} - {m_q \over 4 \pi^2
(x-y)^2}~.
\eea
In order to evaluate Eq. (\ref{ete13}) we substitute the light quark propagator,
we first take the derivatives with respect to $x$ and $y$, and then set
$y=0$. The correlation function receives contributions from the following
three sources:

\begin{itemize}

\item Perturbative contributions.

\item ``Mixed contributions". This contribution takes place when photon
interacts with the quark fields perturbatively and the other quark fields
interact with the QCD vacuum, i.e., they form condensates.

\item Long distance contributions, i.e., photon is radiated at long
distance.

\end{itemize}

The perturbative contribution can be obtained from Eq. (\ref{ete13}) by
replacing on of the propagators by:
\bea
\label{ete16}
S(x-y) = \int d^4z \, S^{free}(x-z) \not\!\!{A} (z)  S^{free}(z-y)~,
\eea
and the  other propagator is chosen as the free quark propagator.

In order to calculate the mixed contribution, it is enough to replace one of
the propagators in Eq. (\ref{ete13}) by Eq. (\ref{ete16}). The
remaining propagator is  replaced by  the quark condensates.

Calculation of long distance contribution proceeds as follows. One of the
quark propagator is replaced by
\bea
\label{nolabel}
S_{\alpha\beta}^{ab} (x-y) = - {1\over 4} \bar{q}^a (x)  \Gamma_j q^b (y)
\left( \Gamma_j \right)_{\alpha\beta}~, \nnb
\eea
where $\Gamma_j = \Big\{ I,~\gamma_\mu,~\gamma_5,~i
\gamma_5\gamma_\mu,~\sigma_{\mu\nu}/\sqrt{2} \Big\}$ are the full set of
Dirac matrices. Since a photon interacts with quark fields at long distance there appears
the matrix elements of nonlocal operators $\bar{q} (x) \Gamma q^\prime(y)$
and $\bar{q} (x) G_{\mu\nu} \Gamma q^\prime(y)$ between the vacuum and photon
state. These are the matrix elements of the photon DA's. The other
remaining propagator is either replaced by the free quark operator, or by the quark condensate.

The matrix elements of the nonlocal operators $\bar{q}\Gamma q^\prime$ and
$\bar{q} G_{\mu\nu} \Gamma q^\prime$ between one photon and the vacuum
states are determined in terms of photon DA's in the following way
\cite{Rte23}:
\bea
\label{ete17}
&&\langle \gamma(q) \ve  \bar q(x) \sigma_{\mu \nu} q(0) \ve  0
\rangle  = -i e_q \bar q q (\varepsilon_\mu q_\nu - \varepsilon_\nu
q_\mu) \int_0^1 du e^{i \bar u qx} \Bigg(\chi \varphi_\gamma(u) +
\frac{x^2}{16} \mathbb{A}  (u) \Bigg) \nnb \\ &&
-\frac{i}{2(qx)}  e_q \qq \Bigg[x_\nu \Bigg(\varepsilon_\mu - q_\mu
\frac{\varepsilon x}{qx}\Bigg) - x_\mu \Bigg(\varepsilon_\nu -
q_\nu \frac{\varepsilon x}{q x}\Bigg) \Bigg] \int_0^1 du e^{i \bar
u q x} h_\gamma(u)
\nnb \\
&&\langle \gamma(q) \ve  \bar q(x) \gamma_\mu q(0) \ve 0 \rangle
= e_q f_{3 \gamma} \Bigg(\varepsilon_\mu - q_\mu \frac{\varepsilon
x}{q x} \Bigg) \int_0^1 du e^{i \bar u q x} \psi^v(u)
\nnb \\
&&\langle \gamma(q) \ve \bar q(x) \gamma_\mu \gamma_5 q(0) \ve 0
\rangle  = - \frac{1}{4} e_q f_{3 \gamma} \epsilon_{\mu \nu \alpha
\beta } \varepsilon^\nu q^\alpha x^\beta \int_0^1 du e^{i \bar u q
x} \psi^a(u)
\nnb \\
&&\langle \gamma(q) \ve \bar q(x) g_s G_{\mu \nu} (v x) q(0) \ve 0
\rangle = -i e_q \qq \Bigg(\varepsilon_\mu q_\nu - \varepsilon_\nu
q_\mu \Bigg) \int {\cal D}\alpha_i e^{i (\alpha_{\bar q} + v
\alpha_g) q x} {\cal S}(\alpha_i)
\nnb \\
&&\langle \gamma(q) \ve \bar q(x) g_s \tilde G_{\mu \nu} i \gamma_5 (v
x) q(0) \ve 0 \rangle = -i e_q \qq \Bigg(\varepsilon_\mu q_\nu -
\varepsilon_\nu q_\mu \Bigg) \int {\cal D}\alpha_i e^{i
(\alpha_{\bar q} + v \alpha_g) q x} \tilde {\cal S}(\alpha_i)
\nnb \\
&&\langle \gamma(q) \ve \bar q(x) g_s \tilde G_{\mu \nu}(v x)
\gamma_\alpha \gamma_5 q(0) \ve 0 \rangle = e_q f_{3 \gamma}
q_\alpha (\varepsilon_\mu q_\nu - \varepsilon_\nu q_\mu) \int {\cal
D}\alpha_i e^{i (\alpha_{\bar q} + v \alpha_g) q x} {\cal
A}(\alpha_i)
\nnb \\
&&\langle \gamma(q) \ve \bar q(x) g_s G_{\mu \nu}(v x) i
\gamma_\alpha q(0) \ve 0 \rangle = e_q f_{3 \gamma} q_\alpha
(\varepsilon_\mu q_\nu - \varepsilon_\nu q_\mu) \int {\cal
D}\alpha_i e^{i (\alpha_{\bar q} + v \alpha_g) q x} {\cal
V}(\alpha_i) \nnb \\ && \langle \gamma(q) \ve \bar q(x)
\sigma_{\alpha \beta} g_s G_{\mu \nu}(v x) q(0) \ve 0 \rangle  =
e_q \qq \Bigg\{
        \Bigg[\Bigg(\varepsilon_\mu - q_\mu \frac{\varepsilon x}{q x}\Bigg)\Bigg(g_{\alpha \nu} -
        \frac{1}{qx} (q_\alpha x_\nu + q_\nu x_\alpha)\Bigg) q_\beta
\nnb \\ \ek
         \Bigg(\varepsilon_\mu - q_\mu \frac{\varepsilon x}{q x}\Bigg)\Bigg(g_{\beta \nu} -
        \frac{1}{qx} (q_\beta x_\nu + q_\nu x_\beta)\Bigg) q_\alpha
\nnb \\ \ek
         \Bigg(\varepsilon_\nu - q_\nu \frac{\varepsilon x}{q x}\Bigg)\Bigg(g_{\alpha \mu} -
        \frac{1}{qx} (q_\alpha x_\mu + q_\mu x_\alpha)\Bigg) q_\beta
\nnb \\ \ar
         \Bigg(\varepsilon_\nu - q_\nu \frac{\varepsilon x}{q.x}\Bigg)\Bigg( g_{\beta \mu} -
        \frac{1}{qx} (q_\beta x_\mu + q_\mu x_\beta)\Bigg) q_\alpha \Bigg]
   \int {\cal D}\alpha_i e^{i (\alpha_{\bar q} + v \alpha_g) qx} {\cal T}_1(\alpha_i)
\nnb \\ \ar
        \Bigg[\Bigg(\varepsilon_\alpha - q_\alpha \frac{\varepsilon x}{qx}\Bigg)
        \Bigg(g_{\mu \beta} - \frac{1}{qx}(q_\mu x_\beta + q_\beta x_\mu)\Bigg) q_\nu
\nnb \\ \ek
         \Bigg(\varepsilon_\alpha - q_\alpha \frac{\varepsilon x}{qx}\Bigg)
        \Bigg(g_{\nu \beta} - \frac{1}{qx}(q_\nu x_\beta + q_\beta x_\nu)\Bigg)  q_\mu
\nnb \\ \ek
         \Bigg(\varepsilon_\beta - q_\beta \frac{\varepsilon x}{qx}\Bigg)
        \Bigg(g_{\mu \alpha} - \frac{1}{qx}(q_\mu x_\alpha + q_\alpha x_\mu)\Bigg) q_\nu
\nnb \\ \ar
         \Bigg(\varepsilon_\beta - q_\beta \frac{\varepsilon x}{qx}\Bigg)
        \Bigg(g_{\nu \alpha} - \frac{1}{qx}(q_\nu x_\alpha + q_\alpha x_\nu) \Bigg) q_\mu
        \Bigg]
    \int {\cal D} \alpha_i e^{i (\alpha_{\bar q} + v \alpha_g) qx} {\cal T}_2(\alpha_i)
\nnb \\ \ar
        \frac{1}{qx} (q_\mu x_\nu - q_\nu x_\mu)
        (\varepsilon_\alpha q_\beta - \varepsilon_\beta q_\alpha)
    \int {\cal D} \alpha_i e^{i (\alpha_{\bar q} + v \alpha_g) qx} {\cal T}_3(\alpha_i)
\nnb \\ \ar
        \frac{1}{qx} (q_\alpha x_\beta - q_\beta x_\alpha)
        (\varepsilon_\mu q_\nu - \varepsilon_\nu q_\mu)
    \int {\cal D} \alpha_i e^{i (\alpha_{\bar q} + v \alpha_g) qx} {\cal T}_4(\alpha_i)
                        \Bigg\},
\eea
where
$\varphi_\gamma(u)$ is the leading twist 2, $\psi^v(u)$,
$\psi^a(u)$, ${\cal A}$ and ${\cal V}$ are the twist 3 and
$h_\gamma(u)$, $\mathbb{A}$, ${\cal T}_i$ ($i=1,~2,~3,~4$) are the
twist 4 photon DA's, respectively and  $\chi$ is the magnetic susceptibility
of the quark fields. The photon DA's are calculated in \cite{Rte23}, and we will
give their explicit forms in the following section.

The measure ${\cal D} \alpha_i$ is defined as
\bea
\label{ete18}
\int {\cal D} \alpha_i = \int_0^1 d \alpha_{\bar q} \int_0^1
d \alpha_q \int_0^1 d \alpha_g \delta(1-\alpha_{\bar q}-
\alpha_q-\alpha_g).\nnb \\
\eea

Carrying out the above--mentioned calculations and separating out the
coefficient of the structure  $\varepsilon^{\prime\prime\beta} q^\nu  g^{\mu\alpha}$  from the
QCD side of the correlation function and equating this result to the
coefficient of the same structure from physical side, the sum rules for the
magnetic moments of the light tensor mesons are obtained. In order to
suppress the  contribution of the higher states and continuum, we apply
double Borel transformation with respect to the variables $p^2$ and
$(p+q)^2$. As a result of the above procedure we obtain:

\bea
\label{ete19}
G_{M_1} (q^2=0) \es {4\over m_T^6 g_T^2} e^{m_T^2/M^2} \Bigg\{
{- 1\over 24} M^2 E_0(x) \Big[ e_{q_1} m_{q_2} \langle \bar{q}_1 q_1 \rangle
\Big( \Bbb{A}(u_0) + 2 (u+u_0)
i_1(h_\gamma)  - 2 \widetilde{i}_1(h_\gamma) \Big) \nnb \\
\ek e_{q_2} m_{q_1} \langle \bar{q}_2 q_2 \rangle \Big( \Bbb{A}(\bar{u}_0) + 2
(\bar{u}+u_0) i_2(h_\gamma)  + \widetilde{i}_2(h_\gamma) \Big) \Big]
\nnb \\
\ar {1\over 32 \pi^2} M^4 E_1(x) (e_{q_1} - e_{q_2}) m_{q_1} m_{q_2}
\Bigg(\gamma_E + \ln {\Lambda^2 \over M^2} \Bigg) \nnb \\
\ar {1\over 48 \pi^2} M^4 E_1(x) (3-4 u_0) (e_{q_1} - e_{q_2}) m_{q_1} m_{q_2}
\nnb \\
\ek {1\over 48} f_{3\gamma} M^4 E_1(x) \Big[e_{q_2} \Big( 8 i_2(\psi_v) -
\psi_a(\bar{u}_0) + (\bar{u}+u_0) (4 \psi_v(\bar{u}_0)
+\psi_a^\prime(\bar{u}_0)) \Big) \nnb \\
\ek e_{q_1} \Big(8 i_1(\psi_v) -
\psi_a(u_0) + (u+u_0) (4 \psi_v(u_0)
-\psi_a^\prime(u_0))\Big)\Big] \nnb \\
\ek {1\over 240 \pi^2} M^6 E_2(x) (5 - 18 u_0) (e_{q_1} - e_{q_2}) \nnb \\
\ar {1\over 32 \pi^2} (e_{q_1} - e_{q_2}) m_{q_1} m_{q_2} \widetilde{j}(s_0)
\nnb \\
\ar {1\over 16 M^2} m_0^2 \Big(e_{q_2} m_{q_2} \langle \bar{q}_1 q_1 \rangle
- e_{q_1} m_{q_1} \langle \bar{q}_2 q_2 \rangle \Big)
\Bigg[ j(s_0) + M^2 \Bigg(2 \gamma_E + \ln {\Lambda^2 \over M^2} \Bigg) \Bigg]
\nnb \\
\ek {1\over 32 \pi^2} M^4 E_1(x) (e_{q_1} - e_{q_2}) m_{q_1} m_{q_2} \nnb \\
\ek {1\over 72} m_0^2 [e_{q_2} \langle \bar{q}_1 q_1 \rangle (m_{q_1} -
3 m_{q_2}) + e_{q_1} \langle \bar{q}_2 q_2 \rangle (3 m_{q_1} -
m_{q_2})] \Bigg\}~,
\eea
where
\bea
\label{nolabel}
i_1(f(u^\prime)) \es \int_{u_0}^1 du^\prime f(u^\prime)~, \nnb \\
\widetilde{i}_1(f(u^\prime)) \es \int_{u_0}^1
du^\prime (u^\prime - u_0) f(u^\prime)~, \nnb \\
i_2(f(u^\prime)) \es \int_{0}^{\bar{u}_0} du^\prime
f(u^\prime)~, \nnb \\
\widetilde{i}_2(f(u^\prime)) \es \int_{0}^{\bar{u}_0} du^\prime
(u^\prime - \bar{u}_0) f(u^\prime)~,\nnb \\
j(s_0) \es \int_0^{s_0} ds \Bigg(\ln {s\over \Lambda^2} e^{-s/M^2}\Bigg)~, \nnb \\
\widetilde{j}(s_0) \es \int_0^{s_0} ds \Bigg( s \ln {s\over \Lambda^2} e^{-s/M^2}
\Bigg)~, \nnb \\
E_n(x) \es 1 - e^{-x} \sum_{k=0}^{n} {x^k\over k!}
= {1\over n!} \int_0^x dx^\prime x^{\prime n} e^{-x^\prime}~,\nnb
\eea
with $x=s_0/M^2$, $s_0$ being the continuum threshold, and the Borel
parameter $M^2$ is defined as:
\bea
\label{nolabel}
M^2 = {M_1^2 M_2^2 \over M_1^2 + M_2^2},
\eea
and
\bea
\label{nolabel1}
u_0={M_1^2\over
M_1^2 + M_2^2}\approx{m_i^2\over
m_i^2 + m_f^2}~, 
\eea
where, $m_i$ and $m_f$ are the mass of the initial and final states.
Remembering that the mass of the initial and final states are the same, therefore it is quite natural to expect that the Borel mass parameters should be very close, hence we
set $M_1^2=M_2^2=2 M^2$ and $u_0=1/2$, which means that the quark and antiquark each carries half of the photon's momentum. Here, we should say that  terms proportional to the  
gluon field strength tensor and quark condensates are small compared to the main nonperturbative contribution
 comes from leading twist distribution amplitudes existing in light cone QCD sum rules approach.
 Therefore, we have omitted numerically ignorable  terms proportional to the gluon field strength tensor from the expression of the magnetic dipole moment
 in Eq. (\ref{ete19}). 

\section{Numerical analysis}

This section encompasses our numerical analysis on the magnetic dipole
moment of the light tensor mesons, $G_{M_1}$. The parameters used in the analysis of the sum rules are as follows:
$\uu(1~GeV) = \dd(1~GeV)= -(0.243)^3~GeV^3$, $\sp(1~GeV) = 0.8
\uu(1~GeV)$, $m_0^2(1~GeV) = (0.8\pm0.2)~GeV^2$ \cite{Rte24},  $m_{f_2} =
(1275 \pm 1.2)~MeV$, $m_{a_2} = (1318.3 \pm 0.6)~MeV$,
 $m_{K_2^\ast}(1430) = (1425.6 \pm 1.5)~MeV$
\cite{Rte25} and  $f_{3 \gamma} = - 0.0039~GeV^2$ \cite{Rte23}. The
magnetic susceptibility is chosen as $\chi(1~GeV)=-(3.15\pm0.3)~GeV^{-2}$
\cite{Rte23}. From sum rules for the  magnetic dipole $G_{M_1}$, it is clear that we also need to know  the decay
constants of the light unflavored and strange tensor mesons. Their values  are taken as
$g_T=0.04$ \cite{Rte18} and $g_T=0.050 \pm 0.002$ \cite{Rte19}, respectively.
The explicit forms of the  photon DA's which are needed in the numerical
calculations are as follows \cite{Rte23}:
\bea
\label{ete20}
\varphi_\gamma(u) \es 6 u \bar u \Big[ 1 + \varphi_2(\mu)
C_2^{\frac{3}{2}}(u - \bar u) \Big]~,
\nnb \\
\psi^v(u) \es 3 [3 (2 u - 1)^2 -1 ]+\frac{3}{64} (15
w^V_\gamma - 5 w^A_\gamma)
                        [3 - 30 (2 u - 1)^2 + 35 (2 u -1)^4]~,
\nnb \\
\psi^a(u) \es [1- (2 u -1)^2] [ 5 (2 u -1)^2 -1 ]
\frac{5}{2}
    \Bigg(1 + \frac{9}{16} w^V_\gamma - \frac{3}{16} w^A_\gamma
    \Bigg)~,
\nnb \\
{\cal A}(\alpha_i) \es 360 \alpha_q \alpha_{\bar q} \alpha_g^2
        \Bigg[ 1 + w^A_\gamma \frac{1}{2} (7 \alpha_g - 3)\Bigg]~,
\nnb \\
{\cal V}(\alpha_i) \es 540 w^V_\gamma (\alpha_q - \alpha_{\bar q})
\alpha_q \alpha_{\bar q}
                \alpha_g^2~,
\nnb \\
h_\gamma(u) \es - 10 (1 + 2 \kappa^+ ) C_2^{\frac{1}{2}}(u
- \bar u)~,
\nnb \\
\mathbb{A}(u) \es 40 u^2 \bar u^2 (3 \kappa - \kappa^+ +1 ) +
        8 (\zeta_2^+ - 3 \zeta_2) [u \bar u (2 + 13 u \bar u) +
                2 u^3 (10 -15 u + 6 u^2) \ln(u) \nnb \\
\ar 2 \bar u^3 (10 - 15 \bar u + 6 \bar u^2)
        \ln(\bar u) ]~,
\nnb \\
{\cal T}_1(\alpha_i) \es -120 (3 \zeta_2 + \zeta_2^+)(\alpha_{\bar
q} - \alpha_q)
        \alpha_{\bar q} \alpha_q \alpha_g~,
\nnb \\
{\cal T}_2(\alpha_i) \es 30 \alpha_g^2 (\alpha_{\bar q} - \alpha_q)
    [(\kappa - \kappa^+) + (\zeta_1 - \zeta_1^+)(1 - 2\alpha_g) +
    \zeta_2 (3 - 4 \alpha_g)]~,
\nnb \\
{\cal T}_3(\alpha_i) \es - 120 (3 \zeta_2 - \zeta_2^+)(\alpha_{\bar
q} -\alpha_q)
        \alpha_{\bar q} \alpha_q \alpha_g~,
\nnb \\
{\cal T}_4(\alpha_i) \es 30 \alpha_g^2 (\alpha_{\bar q} - \alpha_q)
    [(\kappa + \kappa^+) + (\zeta_1 + \zeta_1^+)(1 - 2\alpha_g) +
    \zeta_2 (3 - 4 \alpha_g)]~,\nnb \\
{\cal S}(\alpha_i) \es 30\alpha_g^2\{(\kappa +
\kappa^+)(1-\alpha_g)+(\zeta_1 + \zeta_1^+)(1 - \alpha_g)(1 -
2\alpha_g)\nnb \\
\ar\zeta_2
[3 (\alpha_{\bar q} - \alpha_q)^2-\alpha_g(1 - \alpha_g)]\}~,\nnb \\
\tilde {\cal S}(\alpha_i) \es-30\alpha_g^2\{(\kappa -
\kappa^+)(1-\alpha_g)+(\zeta_1 - \zeta_1^+)(1 - \alpha_g)(1 -
2\alpha_g)\nnb \\
\ar\zeta_2 [3 (\alpha_{\bar q} -
\alpha_q)^2-\alpha_g(1 - \alpha_g)]\}.
\eea
The constants entering  the above DA's are borrowed from
\cite{Rte05} whose values are $\varphi_2(1~GeV) = 0$,
$w^V_\gamma = 3.8 \pm 1.8$, $w^A_\gamma = -2.1 \pm 1.0$,
$\kappa = 0.2$, $\kappa^+ = 0$, $\zeta_1 = 0.4$, $\zeta_2 = 0.3$,
$\zeta_1^+ = 0$ and $\zeta_2^+ = 0$.

The sum rules for the magnetic dipole moment of the tensor mesons contain two more
auxiliary parameters, namely, continuum threshold $s_0$ and Borel parameter
$M^2$. The continuum threshold is not completely arbitrary and it is related
to the energy of the excited states. From our analysis we observe that
$G_{M_1}$ is practically independent of $s_0$ in the interval $(m_T + 0.3)^2 \le
s_0 \le (m_T + 0.6)^2$. Note that, this region of $s_0$ practically coincides with the  region of $s_0$ used in analysis of the mass sum rules for the light tensor mesons (for details see \cite{baganb}). The working region for the Borel parameter
$M^2$ is obtained by the following procedure: The upper limit of $M^2$ is
determined by requiring that the series of the light cone expansion with
increasing twist should be convergent. The lower limit of $M^2$ is obtained
by the fact that the contribution of the higher states and continuum to the
correlation function should be small enough. The above requirements restrict
the working region of the Borel parameter to $1~GeV^2 \le M^2 \le 3~GeV^2$.

Using the photon DA's and the working regions of $s_0$ and $M^2$, we obtain
the values of the magnetic dipole moments of the light tensor mesons for both the
charged and neutral cases, which are presented in Table 1.

\begin{table}[h]

\renewcommand{\arraystretch}{1.3}
\addtolength{\arraycolsep}{-0.5pt}
\small
$$
\begin{array}{|l|r@{\pm}l|}
\hline \hline
\mbox{\rm Tensor Meson} &
\multicolumn{2}{c|}{G_{M_1}~(e/2 m_T)} \\ \hline
f_2^\pm              &   2.1  &0.5   \\
f_2^0                &   0.0  &0.0   \\
a_2^\pm              &   1.88 &0.4   \\
a_2^0                &   0.0  &0.0   \\
K_2^{\ast \pm}(1430) &   0.75 &0.08  \\
K_2^{\ast 0}(1430)   &   0.076&0.008 \\
\hline \hline

\end{array}
$$
\caption{Magnetic dipole moments of the tensor mesons in units of $e/2 m_T$.}
\renewcommand{\arraystretch}{1}
\addtolength{\arraycolsep}{-1.0pt}

\end{table}

The values of the magnetic dipole moments presented in Table 1 are in units of
$e/2 m_T$, and the quoted errors are due to the uncertainties in the input
parameters, that is, the parameters entering the photon DA's, as well as,
the working region for the continuum threshold $s_0$, and the Borel
parameter $M^2$. We see from the table that the value of the magnetic dipole
moments for the charged tensor mesons are considerably different from zero,
which can be attributed to the response of the tensor mesons to an external
magnetic field. The magnetic dipole moments for the neutral and unflavored
light tensor mesons are very close to zero, while it has a nonzero small value
for the $K_2^{\ast 0}(1430)$ tensor meson.

Finally, let us compare our results on magnetic moments of light tensor
mesons with the existing lattice QCD results \cite{Rte12}. For example, the magnetic
moment of $K_2^{\ast \pm}(1430)$ mesons in lattice QCD change between $\pm 0.5$ and
$\pm 0.8$, depending on effective mass of the $\pi$ meson and  our analysis
predicts $\pm(0.75 \pm 0.08)$. We see that our prediction is in good
agreement with the lattice results, especially for the large effective
$\pi$ meson mass case. Our result on the magnetic moment of the neutral
$K_2^{\ast 0}(1430)$ meson is $\pm(0.076 \pm 0.008)$, which is slightly
higher compared to the lattice QCD prediction $\pm 0.05$. Note that in calculations, the SU(2) flavor symmetry is implied. The nonzero value of the magnetic moment of $K_2^{\ast 0}(1430)$ is due to the SU(3) flavor symmetry breaking.

In summary, the magnetic dipole moment of the light tensor mesons have been
calculated in the framework of the LCSR using the photon DA's. We observe
that the magnetic moments of charged light tensor mesons are
practically 2.5--3 times larger compared to that of the magnetic moments of
the charged strange tensor mesons. The magnetic moment of the neutral strange
tensor meson is also nonzero, but its value is small.
However, the values of the magnetic dipole moments of the light, neutral,
unflavored tensor mesons are very close to zero.

\section{Acknowledgment}
We thank A. Ozpineci for his useful discussions.

\end{document}